\documentclass[12pt,
aps,
 pra,reprint,
 superscriptaddress,
longbibliography
]{revtex4-1}
\usepackage{dcolumn}% Align table columns on decimal point
\usepackage{bm}% bold math
\usepackage%
{graphicx}
\usepackage{xfrac}
\usepackage{extdash}
\usepackage{epsfig}
\usepackage{isomath}
\usepackage{textcomp}
\usepackage{bm}
\usepackage[english]{babel}
\usepackage{todonotes}
\usepackage{color,soul}
\usepackage{subcaption}

\usepackage{times}

\DeclareFontFamily{U}{euc}{}
\DeclareFontShape{U}{euc}{m}{n}{<-6>eurm5<6-8>eurm7<8->eurm10}{}% 
\DeclareSymbolFont{AMSc}{U}{euc}{m}{n} 
\DeclareMathSymbol{\umu}{\mathord}{AMSc}{"16} 

\renewcommand{\vec}[1]{\boldsymbol{#1}}
\newcommand{\ensuretext}[1]{\ensuremath{\text{#1}}}

\newcommand{\unit}[1]{\ensuretext{\textrm{\,}}\ensuremath{\mathrm{#1}}}
\newcommand{\eV}{\mathrm{eV}}
\newcommand{\MeV}{\mathrm{M}\eV}
\newcommand{\keV}{\mathrm{k}\eV}

\newcommand{\Mum}{\ensuremath{\umu}\ensuremath{\mathrm{m}}}
\newcommand{\mum}{\textrm{\,\ensuremath{\mathrm{\Mum}}}}
\newcommand{\eqref}[1]{(\ref{#1})}

\begin{document}

\title{Probing the dynamics of solid density micro-wire targets after ultra-intense laser irradiation using a free-electron laser} 

\author{Thomas Kluge}
\affiliation{Helmholtz-Zentrum Dresden-Rossendorf, Bautzner Landstra\ss e 400, 01328, Dresden, Germany}
\email{t.kluge@hzdr.de} 
\author{Michael Bussmann}
\affiliation{Helmholtz-Zentrum Dresden-Rossendorf, Bautzner Landstra\ss e 400, 01328, Dresden, Germany}
\author{Eric Galtier}
\affiliation{SLAC National Accelerator Laboratory, 2575 Sand Hill Rd, Menlo Park, CA 94025, USA}
\author{Siegfried Glenzer}
\affiliation{SLAC National Accelerator Laboratory, 2575 Sand Hill Rd, Menlo Park, CA 94025, USA}
\author{J\"org Grenzer}
\affiliation{Helmholtz-Zentrum Dresden-Rossendorf, Bautzner Landstra\ss e 400, 01328, Dresden, Germany}
\author{Christian Gutt}
\affiliation{Universit\"at Siegen, Adolf-Reichwein-Stra\ss e 2, 57068 Siegen, Germany}
\author{Nicholas J. Hartley}
\affiliation{SLAC National Accelerator Laboratory, 2575 Sand Hill Rd, Menlo Park, CA 94025, USA}
\author{Lingen Huang}
\affiliation{Helmholtz-Zentrum Dresden-Rossendorf, Bautzner Landstra\ss e 400, 01328, Dresden, Germany}
\author{Alejandro Laso Garcia}
\affiliation{Helmholtz-Zentrum Dresden-Rossendorf, Bautzner Landstra\ss e 400, 01328, Dresden, Germany}
\author{Hae Ja Lee}
\affiliation{SLAC National Accelerator Laboratory, 2575 Sand Hill Rd, Menlo Park, CA 94025, USA}
\author{Emma E. McBride}
\affiliation{European XFEL, Holzkoppel 4, 22869 Schenefeld, Germany}
\affiliation{SLAC National Accelerator Laboratory, 2575 Sand Hill Rd, Menlo Park, CA 94025, USA}
\author{Josefine Metzkes-Ng}
\affiliation{Helmholtz-Zentrum Dresden-Rossendorf, Bautzner Landstra\ss e 400, 01328, Dresden, Germany}
\author{Motoaki Nakatsutsumi}
\affiliation{European XFEL, Holzkoppel 4, 22869 Schenefeld, Germany}
\author{Inhyuk Nam}
\affiliation{SLAC National Accelerator Laboratory, 2575 Sand Hill Rd, Menlo Park, CA 94025, USA}
\affiliation{now with Pohang Accelerator Laboratory, Pohang, Gyeongbuk 37673, KR}
\author{Alexander Pelka}
\affiliation{Helmholtz-Zentrum Dresden-Rossendorf, Bautzner Landstra\ss e 400, 01328, Dresden, Germany}
\author{Irene Prencipe}
\affiliation{Helmholtz-Zentrum Dresden-Rossendorf, Bautzner Landstra\ss e 400, 01328, Dresden, Germany}
\author{Lisa Randolph}
\affiliation{Universit\"at Siegen, Adolf-Reichwein-Stra\ss e 2, 57068 Siegen, Germany}
\affiliation{now with European XFEL, Holzkoppel 4, 22869 Schenefeld, Germany}
\author{Martin Rehwald}
\affiliation{Helmholtz-Zentrum Dresden-Rossendorf, Bautzner Landstra\ss e 400, 01328, Dresden, Germany}
\affiliation{Technical University Dresden,01069 Dresden, Germany}
\author{Christian R\"odel}
\affiliation{SLAC National Accelerator Laboratory, 2575 Sand Hill Rd, Menlo Park, CA 94025, USA}
\affiliation{Friedrich-Schiller-Universit\"at, Max-Wien-Platz 1, 07743 Jena, Germany}
\author{Melanie R\"odel}
\affiliation{Helmholtz-Zentrum Dresden-Rossendorf, Bautzner Landstra\ss e 400, 01328, Dresden, Germany}
\author{Toma Toncian}
\affiliation{Helmholtz-Zentrum Dresden-Rossendorf, Bautzner Landstra\ss e 400, 01328, Dresden, Germany}
\author{Long Yang}
\affiliation{Helmholtz-Zentrum Dresden-Rossendorf, Bautzner Landstra\ss e 400, 01328, Dresden, Germany}
\author{Karl Zeil}
\affiliation{Helmholtz-Zentrum Dresden-Rossendorf, Bautzner Landstra\ss e 400, 01328, Dresden, Germany}
%\author{Uwe H\"ubner}
%\affiliation{Leibniz Institute of Photonic Technology, Albert-Einstein-Stra\ss e 9, 07745 Jena, Germany}
\author{Ulrich Schramm}
\author{Thomas E. Cowan}
\affiliation{Helmholtz-Zentrum Dresden-Rossendorf, Bautzner Landstra\ss e 400, 01328, Dresden, Germany}
\affiliation{Technical University Dresden,01069 Dresden, Germany}

\date{\today}

\begin{abstract}
In this paper, we present an experiment that explores the plasma dynamics of a $7\unit{\mum}$ diameter carbon wire after being irradiated with a near-relativistic-intensity short pulse laser. Using an X-ray Free Electron Laser pulse to measure the small angle X-ray scattering signal, we observe that the scattering surface is bent and prone to instability over tens of picoseconds. The dynamics of this process are consistent with the presence of a sharp, propagating shock front inside the wire, moving at a speed close to the hole boring velocity.

\end{abstract}
\maketitle 

\section*{Introdction}
Modern laser technology allows the generation of ultra-intense optical light, compressed to a few femtoseconds and focused to a few micron spot sizes leading to intensities exceeding the threshold for relativistic electron motion of $\sim 10^{18}\unit{W/cm}^2$. 
When a laser pulse is focused on a solid surface at these intensities, the atoms at the surface are ionized and a dense plasma is created. 
The laser field interacts with electrons within a skin depth of a few tens of nanometers and can compress the surface and accelerate electron and ions to several $\unit{\MeV}$. 
The interaction of such high intensity lasers with solids has attracted considerable interest with the prospect of creating bright and compact particle accelerators~\cite{Dalui2016,*Macchi2013,*Albert2021} and gamma-ray sources~\cite{Stark2015}. \\
Several physical processes occur before or after the laser pulse on a picosecond to nanosecond time scale leading for example to thermalization, diffusion, compression and ablation\cite{Craxton2015}. 
These processes generally take place in the regime of coupled plasmas, where atomic-scale collisions still play a role. 
Processes on the sub-picosecond or even sub-femtosecond scale on the other hand usually are collective, giving rise to complex phenomena and waves. 
One important aspect is the generation of a multitude of instabilities, influencing for example the coupling of the laser to the plasma, particle transport, and heating of the target. 
\begin{figure*}
\centering
    \begin{subfigure}{0.74\linewidth}
      \includegraphics[width=\textwidth]{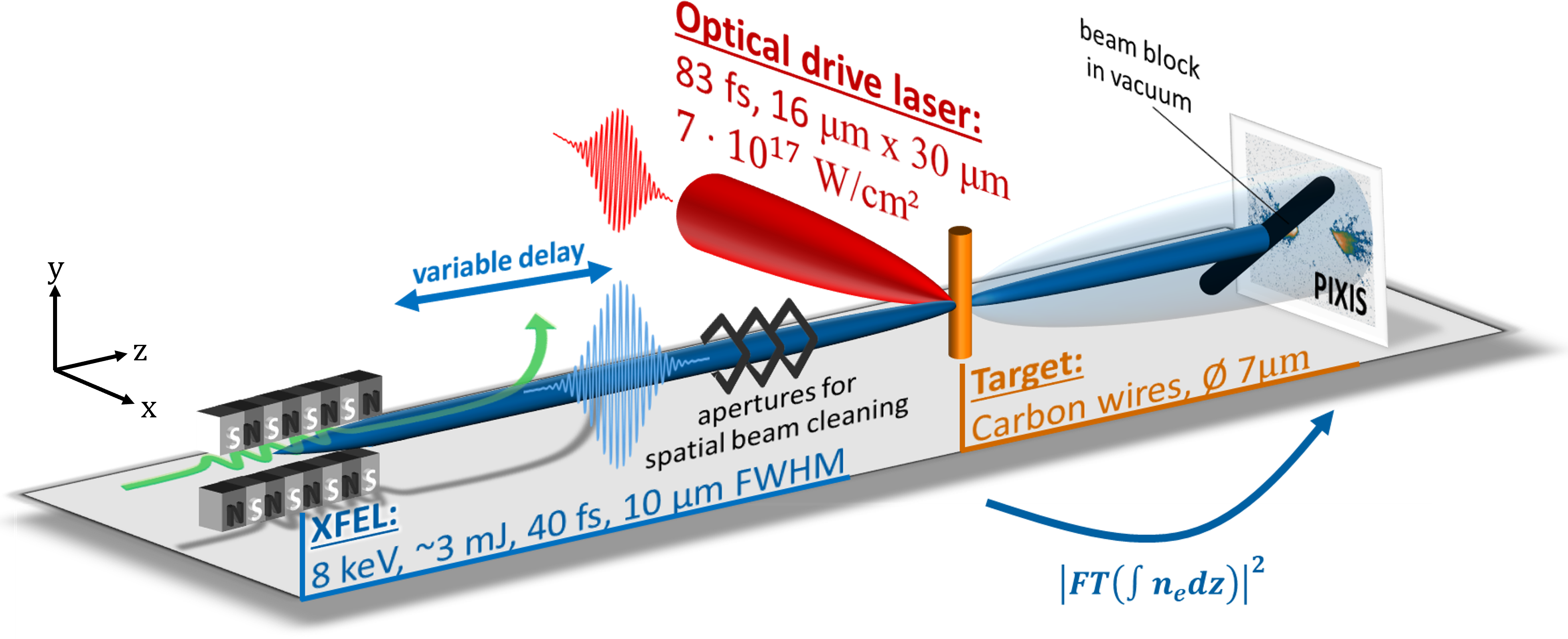}
      \caption{Setup}
      \label{fig:setup:setup}
    \end{subfigure}
    \begin{subfigure}{0.24\linewidth}
      \includegraphics[width=\textwidth]{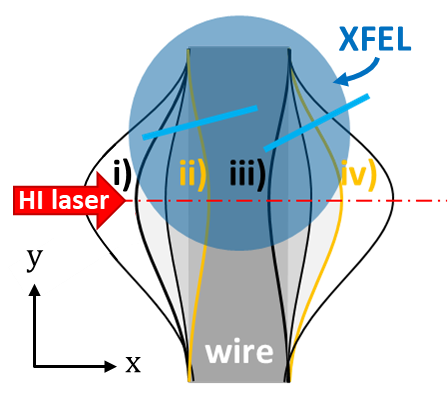}
      \caption{Effect of XFEL vertical offset}
      \label{fig:setup:angles_explained}
    \end{subfigure}
  \caption{\textbf{(a)} Sketch of the setup of the experiment, not to scale. The high intensity short pulse pump laser irradiates the wire at normal incidence, an XFEL pulse probes the plasma under $90^\circ$ incidence. The detector was positioned at a distance to the target of $1.4\unit{m}$. The XFEL was vertically off-centered, effecting the signal: \textbf{(b)} The XFEL scattering streak (blue line) is assumed to stand perpendicular on the scattering surface of the wire. The scattering surface can be (i) the forward ablated surface, (ii) the hole boring (shock) front, (iii) the rear surface rare faction wave, (iv) the ablated rear surface, or a mechanically bent foil. Due to the measured orientation of the streak (see Figs.~\ref{fig:delay_scan},~\ref{fig:I_scan}, i), iii) can be ruled out. }
  \label{fig:setup}
\end{figure*}
So far, such surface plasma dynamics have been widely investigated using femtosecond optical laser pulses~\cite{Sokolowski-Tinten1998}, whereby various surface-sensitive methods based on optical shadowgraphy\cite{martin},  interferometry~\cite{Geindre1994, *Bocoum2015}, and spectroscopy~\cite{Mondal2010,Malvache2013,Hornung2021} have been applied. 
However, optical probes can only penetrate the plasma up to their critical density, i.e. a fraction of the solid density, and so cannot directly give access to the dynamics inside the solid density region. 

X-ray diffraction can be employed to resolve physical processes above the critical density, such as for example non-thermal melting~\cite{Siders1999}, coherent lattice vibrations~\cite{Sokolowski-Tinten2003}, and ultrafast phase transitions~\cite{McBride2019}. 
%Ultrafast deformation within a thin metal film has been studied using laser-driven soft x-ray sources~\cite{Tobey2007}. 
Recently, X-ray pulses from X-ray free electron lasers (XFEL) have been applied to investigate laser produced plasmas with femtosecond temporal resolution. %~\cite{Fletcher2015}. 
Small-angle femtosecond X-ray Scattering (SAXS) has revealed density gradients of expanding solid-density plasmas with nanometer spatial and few femtosecond temporal resolution~\cite{Kluge2014,Gorkhover2016,Kluge2018}, transient nano-jet emission from grating surfaces\cite{Kluge2018}, as well as ultra-fast heating and ionization~\cite{Kluge2016,*Kluge2017,*Gaus2021}. \\

Here we use SAXS to study the dynamics inside solid density wires that turn into plasmas after high-intensity short-pulse laser irradiation. 
Our experimental data shows that the scattering surface is bent by a few microns and prone to instability development also on the scale of a few microns over the course of tens of picoseconds. 
The dynamics of this process are consistent with a sharp, propagating compression (shock) front inside the wire. 
With the novel SAXS method we measure the tilt angle of this front and observe that it is moving forward with a speed close to the hole boring (HB) velocity of up to $0.1\unit{\mum/ps}$. 
%The SAXS method employed in the present paper can be used to investigate nanometer-sized dynamics of hot dense plasmas with femtosecond XFEL pulses. 

\section*{Experimental setup}
We report on an experiment probing the plasma dynamics of a $7\mum$ diameter carbon wire after the irradiation with the near-relativistic high-intensity (HI) Titanium:Sapphire short-pulse near-infrared (wavelength $800\unit{nm}$) laser at the Matter in Extreme Conditions (MEC) endstation at the Linear Coherent Light Source (LCLS) XFEL. 
%probing at earlier times (even during the laser pulse itself) and at smaller scales down to few nm or tens of nm by optimizing the laser and target properties in the future. 
%\cite{Kluge2014}\cite{Kluge2015}\cite{Kluge2016}\cite{Kluge2017}\cite{Kluge2018}\cite{Kluge2018a}\cite{Gaus2021}\cite{zastrau2021}\cite{lasogarcia2021}
\begin{figure*}
\centering
  \includegraphics[width=\linewidth]{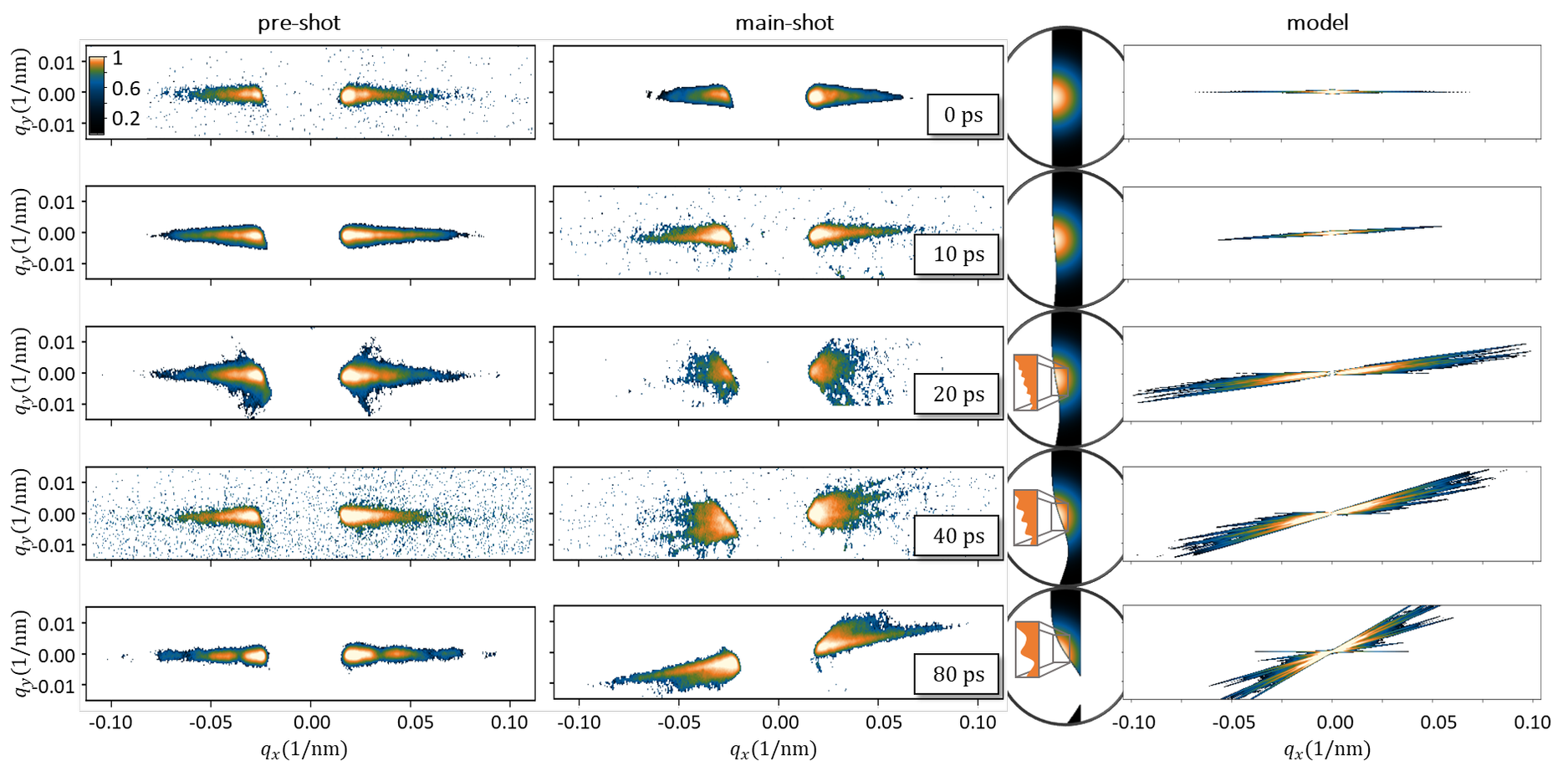}
\caption{Normalized scattering patterns for different XFEL probe delays (probe comes after pump) at an intensity of $5\cdot 10^{17} \unit{W/cm}^2$. The right column shows example scattering patterns based on a hole boring model described in \eqref{eqn:v_HB} and Eqn.~\eqref{eqn:model}: The circles depict the example density distribution in real space (multiplied with the XFEL intensity), the right column shows the respective Fourier absolute square signal, for an $A$ in Eqn.~\eqref{eqn:model} of $0\unit{\mum}$, $1\unit{\mum}$, $2\unit{\mum}$, $4\mum$, $8\unit{\mum}$ (top to bottom). For the cases with delay $20\unit{ps}$, $40\unit{ps}$, $80\unit{ps}$ we added a sinosoidal surface structure with $12.5\unit{nm}$, $25\unit{nm}$, $50\unit{nm}$ amplitude and $1\unit{\mum}$, $2\unit{\mum}$, $3\unit{\mum}$ period, respectively. }
  \label{fig:delay_scan}
\end{figure*}
The HI laser pulse duration was $83\unit{fs}$ and the spot size at FWHM was $w_x\approx 30\unit{\mum}$ and $w_{y}\approx 16\unit{\mum}$ in horizontal and vertical direction, respectively. 
The laser pulse energy was measured to be up to $1\unit{J}$ before compression. 
With $46\% $ transmission to the target and $22\%$ of the energy being in the FWHM area, the peak intensity on target can be calculated to approx. $5\cdot 10^{17}\unit{W/cm}^2$, equivalent to a normalized laser strength amplitude of $a_0\cong 0.5$. 
The XFEL pulse had a diameter of $5-10\unit{\mum}$ and pulse duration of $40\unit{fs}$, both at FWHM. 
The X-ray photon energy was $8\unit{\keV}$, with approximately $N_0=10^{11}$ photons per pulse (the exact number on target varies from shot to shot due to different absorbers). 
Temporal synchronization between the HI short-pulse laser and LCLS XFEL pulse was achieved with a precision of $120\unit{fs}$ rms. \\
%Our samples were $7\unit{\mum}$ thick cylindrical carbon fiber wires. 
The carbon wire targets were irradiated under normal incidence by both the pump and the probe pulse, which were also perpendicular to each other, see Fig.~\ref{fig:setup}. 
For reasons detailed below, the XFEL axis was vertically offset by approximate $20\unit{\mum}$ along the wire towards the top. 

In the following, we present a series of shots on these wires, varying the pump laser intensity between 50\% and 100\% (i.e. between $3.5\cdot 10^{17}\unit{W/cm}^2$ and $7\cdot 10^{17}\unit{W/cm}^2$), and the probe delay $\Delta t$ between $0$ and $80\unit{ps}$. 
We measure the small angle X-ray scattering (SAXS) signal with an PIXIS2048 X-ray detector with $13\unit{\mum}$ pixel size X-ray that was positioned $1.4~\unit{m}$ behind the target. 
In this configuration, the signal is given in Born approximation (i.e. phase contrast only) by the Fourier transform of the electron density, $I(\vec{q})\propto \left|FT(n(\vec{r}))\right|$, where $q=\left|\vec{k}_0-\vec{k}'\right|)$ is the scattering vector defined by the difference between the incoming XFEL wave vector $\vec{k}_0$ and the scattered wave vector $\vec{k}'$. 
Here, we only expect two streaks perpendicular to the wire surface, the interference pattern of scattered waves from the wire front and rear surface (i.e. $q=\lambda_{XFEL}/7\unit{\mum}\approx 2\cdot 10^{-2}\unit{mrad}$) cannot be resolved with the detector resolution ($q=2\pi/13\unit{13}\approx 1\cdot 10^{-3}\unit{mrad}$). 
The setup we employ here enables the study of a multitude of relevant physics following the relativistic laser irradiation~\cite{Kluge2014,Kluge2017}, through the analysis of the change of the scattering streak's signal intensity and slope, orientation, and structure, as discussed in the following. 

\section*{SAXS Method}
First, we would like to extract the temporal evolution of the plasma expansion, as we demonstrated before on grating targets\cite{Kluge2018}. 
Similar to the Debye-Waller analysis for thermal motion of particles, we can replace the displacement of scatterers due to thermal motion with the displacement of the plasma due to non-thermal melt/expansion into vacuum. 
The expansion scale $\sigma$ can then simply be derived from the exponential roll-off of the scattering signal at large scattering angles \cite{Kluge2018,Gorkhover2016} 
\begin{equation}
    I(q)\propto q^{-r} \exp(-q^2 \sigma^2)
    \label{eqn:Debye}
\end{equation} 
where $r$ depends on the geometry of the sample. 
The exponent $r$ is expected for flat surfaces to be in the range of $2-4$, e.g. $r=2$ for a cuboid, $r=3$ for a cylinder and $r=4$ for a sphere\cite{guinier1955}, or $r=3-4$ for a fractally rough surface\cite{Benedetti1993}. 
The exponential term is the Debye-Waller factor. 
\begin{figure*}
\centering
  \includegraphics[width=\linewidth]{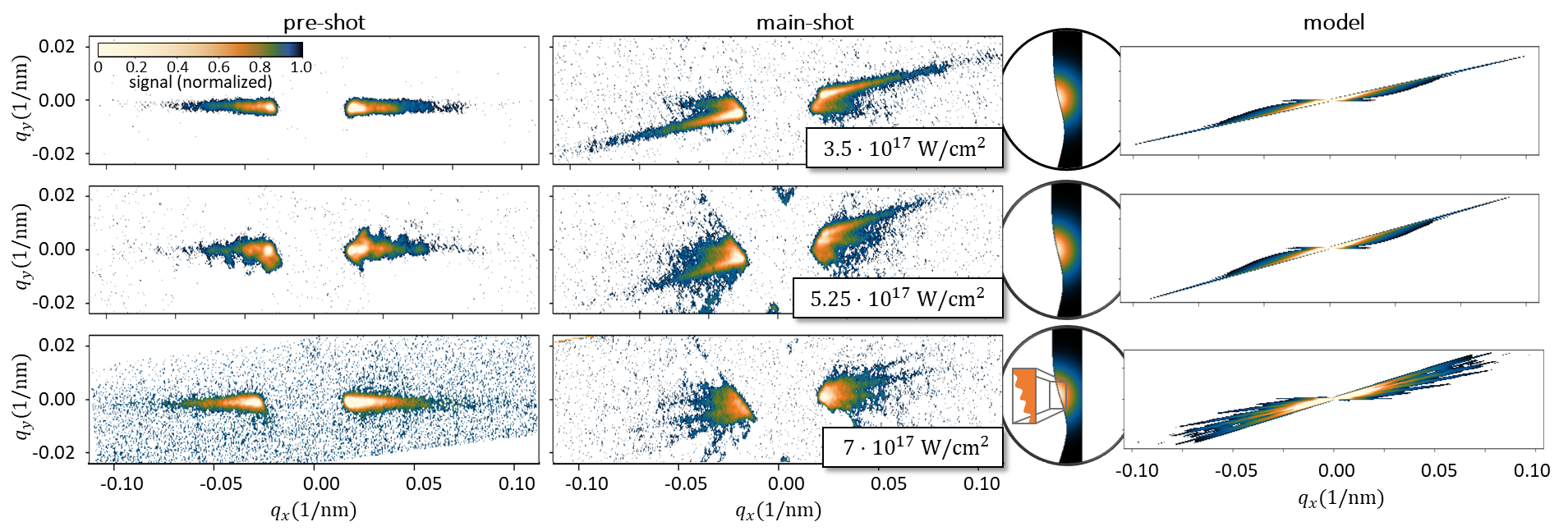}
\caption{Normalized scattering patterns for different laser intensities. XFEL probe delay was set to $40 \unit{ps}$. The right column shows example scattering patterns based on a hole boring model described in \eqref{eqn:v_HB} and Eqn.~\eqref{eqn:model}: The circles depict the example density distribution in real space (multiplied with the XFEL intensity), the right column shows the respective Fourier absolute square signal, for $A=2.8\unit{\mum}$ (top), $3.5\unit{\mum}$ (middle), and $4.0\unit{\mum}$ (bottom) in Eqn.~\eqref{eqn:model}. As in Fig.~\ref{fig:delay_scan_profiles}, for the case at full intensity we added a sinusoidal surface with $25\unit{nm}$ amplitude and $2\unit{\mum}$ period. (!!! add colorbar !!!)}
  \label{fig:I_scan}
\end{figure*}

Secondly, we can expect the scattering streak of the wire to tilt if the wire is being deformed by the laser. 
As depicted in Fig.~\ref{fig:setup:angles_explained}, this can in principle happen due to (i) ablation into vacuum at the front following the thermal pressure set up by laser accelerated electrons or the bulk electrons heated by return current collisions\cite{Kluge2018a,Mora2003}; (ii) a traveling compression (hole boring or shock) front into the bulk launched by the laser pressure\cite{Robinson2009,Fiuza2012a}; (iii) a rarefaction (shock) wave into the wire at the rear surface \cite{Bulgakova1999}; (iv) ablation into vacuum at the rear\cite{Mora2003}. \\
To determine the relevant mechanisms in the present setup, we employ simulations already published for similar conditions (cp. Fig. 3 of \cite{Gaus2021}). 
The laser and XFEL parameters are the same while only the target material differs, being Si in the simulation instead of the carbon used here. 
However, the effective scattering length densities $\int n_e dz$ are quite similar (for silicon it is approx. $2/3$ of that of carbon) and the simulations therefore describe also the present case qualitatively. 
As can be seen in these simulations, over the course of only a few picoseconds the front surface is fully ablated (i.e. order of $\unit{\mum}$ scalelength). 
Such a diluted surface would not be possible to be detected due to the corresponding Debye-Waller factor that would be close to zero and would suppress the streak signal already at small $q$, effectively making the streak vanish. 
Simultaneously, a compression shock wave is launched by the laser pressure, which continues to travel through the target over 10s of picoseconds. 
The shock front velocity is between $0.05\unit{\mum/ps}$ (hydro simulation) and $0.1\unit{\mum/ps}$ (simulation), i.e. around the hole boring velocity  \begin{equation}
    v_{HB}=c\frac{\Xi}{1+\Xi}\cong 0.09\unit{\mum/ps}
    \label{eqn:v_HB}
\end{equation} where the piston parameter for a fully ionized plasma is $\Xi \cong a_0/\sqrt{0.5 m_p n_e/m_e n_c}$, $m_i$ and $m_e$ are the ion and electron mass, respectively, and $n_c=\varepsilon_0 m_e \omega_0^2/e^2$ is the critical density\cite{Robinson2009}. 
The HB velocity in carbon for the present case is computed to an only slightly larger value of $v_{HB}=0.1\unit{\mum/ps}$, hence we can expect the compression wave to break out of the rear surface of the $7\unit{um}$ thick carbon wire after approx. $70\unit{ps}$. \\
Additionally, briefly after the laser irradiation some high energy electrons accelerated by the laser penetrate the full target ballistically, which excites a return current and consequently bulk target heating to some few $eV$. 
Hence, the rear surface expands/ablates slightly due to the thermal pressure. 
Both features, the shock front and the ablating rear surface, as well as the rarefaction wave countering the rear surface expansion, could in principle lead to a signal in form of a streak on our SAXS detector % , and a tilt of the scattering streak 
depending on their exact sharpness. 

Finally, instabilities in the shock front or the rear surface, respectively, can lead to a break-up of the single wire scattering line into multiple lines, which in the case of a sinusoidal surface modulation would each reflect a certain orientation of the target normal at the point of inflection. 
Such instabilites can be, among others, Rayleigh-Taylor like instabilities leading to a surface rippling at relativistic intensities\cite{Palmer2012,Sgattoni2015,Kluge2015}, filamentation of filamentation-two-stream instabilities in the target \cite{Bret2004,Metzkes2014a} or at the target rear surface\cite{Gode2017}, Weibel-like instabilities occurring during the plasma propagation in vacuum after laser acceleration\cite{Quinn2012}, or the Z-pinch sausage (m=0) instability (especially at later times after few ps)\cite{Beg2004}. 
The instabilities lead to a growth of initially small density fluctuations to form ripples even on an initially flat target surface. 

%Optical methods struggle with the observation of above dynamics, e.g. due to low but still overcritical plasma corona quickly developing around the target. 
%X-rays, however can easily penetrate the optically opaque plasma. 
In the regime relevant for this work, the dominant source of small angle X-ray scattering is coherent elastic Thomson scattering from electrons, but there is also a relevant diffractive contribution by bound-free transition absorption. 
The latter depends sensitively on the plasma ionization degree, i.e. the plasma temperature. 
%In the SAXS geometry the scattering pattern is then obtained in kinetic and Born approximation by the absolute square of the exit wave Fourier transform\cite{O.Glatter1982}. 
\begin{figure*}
\centering
  \includegraphics[width=\linewidth]{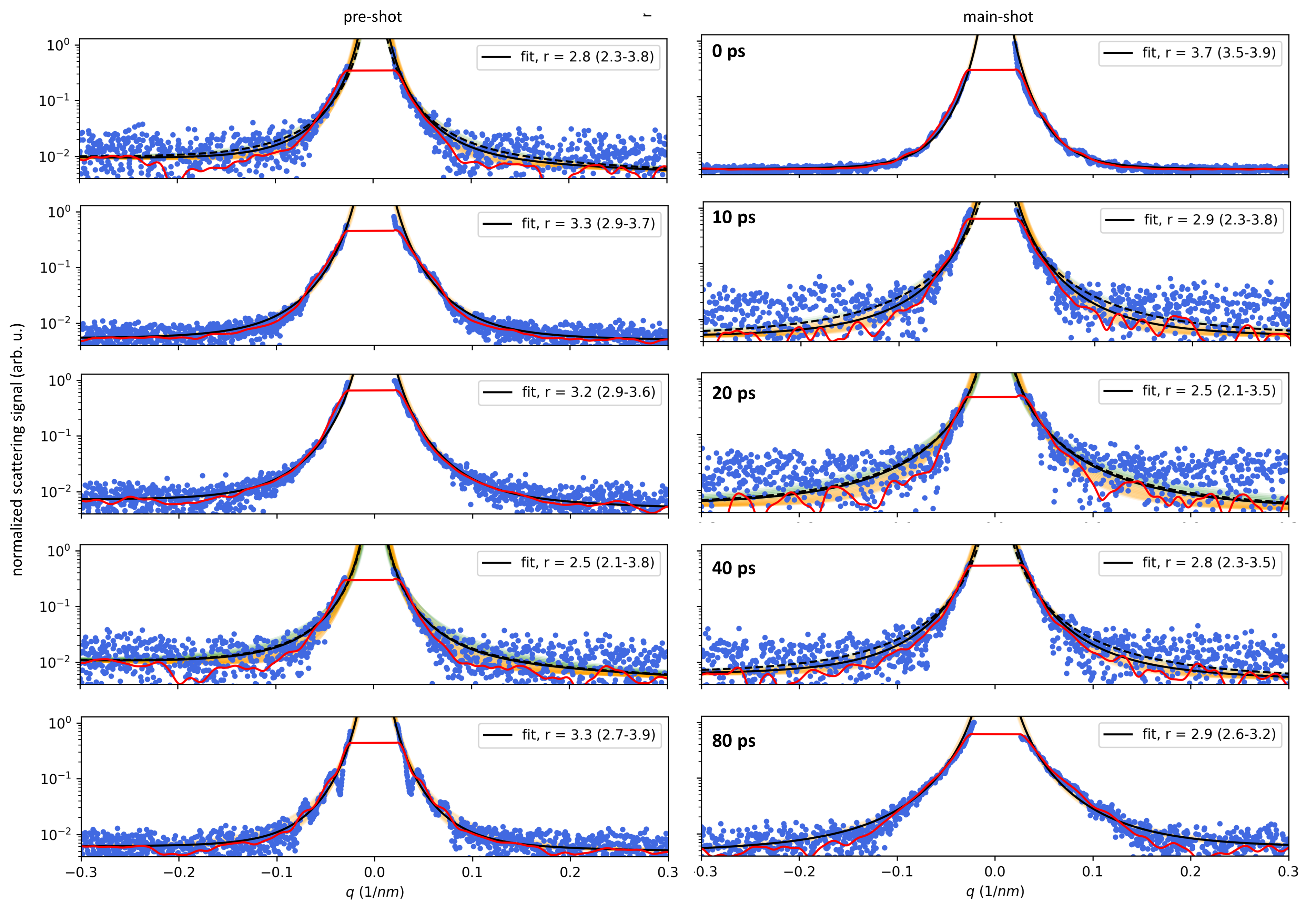}
\caption{Streak profiles for the scattering patterns shown in Fig.~\ref{fig:delay_scan}. The red line is the rolling average over $15\unit{px}$, the black line shows the best fit with the fitted exponent $r$ and its confidence interval ($1\sigma$) given in the respective legend. The black dashed line shows the best fit with $r<2.5$ to demonstrate how insensitive the fits are with respect to $r$. The shaded area is the error of the fit.}
  \label{fig:delay_scan_profiles}
\end{figure*}

\section*{Results}
The scattering patterns recorded in this experiment are shown in  Fig.~\ref{fig:delay_scan} as a function of delay at full laser intensity, and in Fig.~\ref{fig:I_scan} as function of intensity at $40\unit{ps}$ probe delay. 
For each shot we took two XFEL-only pre-shots as a reference, that we averaged for better statistics. 
We used this to determine the undriven target normal orientation and verify the surface quality and XFEL overlap. 
We also took an XFEL-only post-shot after the combined laser-XFEL main-shot, in order to be able to subtract in the pre- and main-shot the parasitic XFEL scattering on the beamstop and other optical components in the beamline. 
The figures show the background subtracted and normalized signal. 

\begin{figure*}
\centering
  \includegraphics[width=\linewidth]{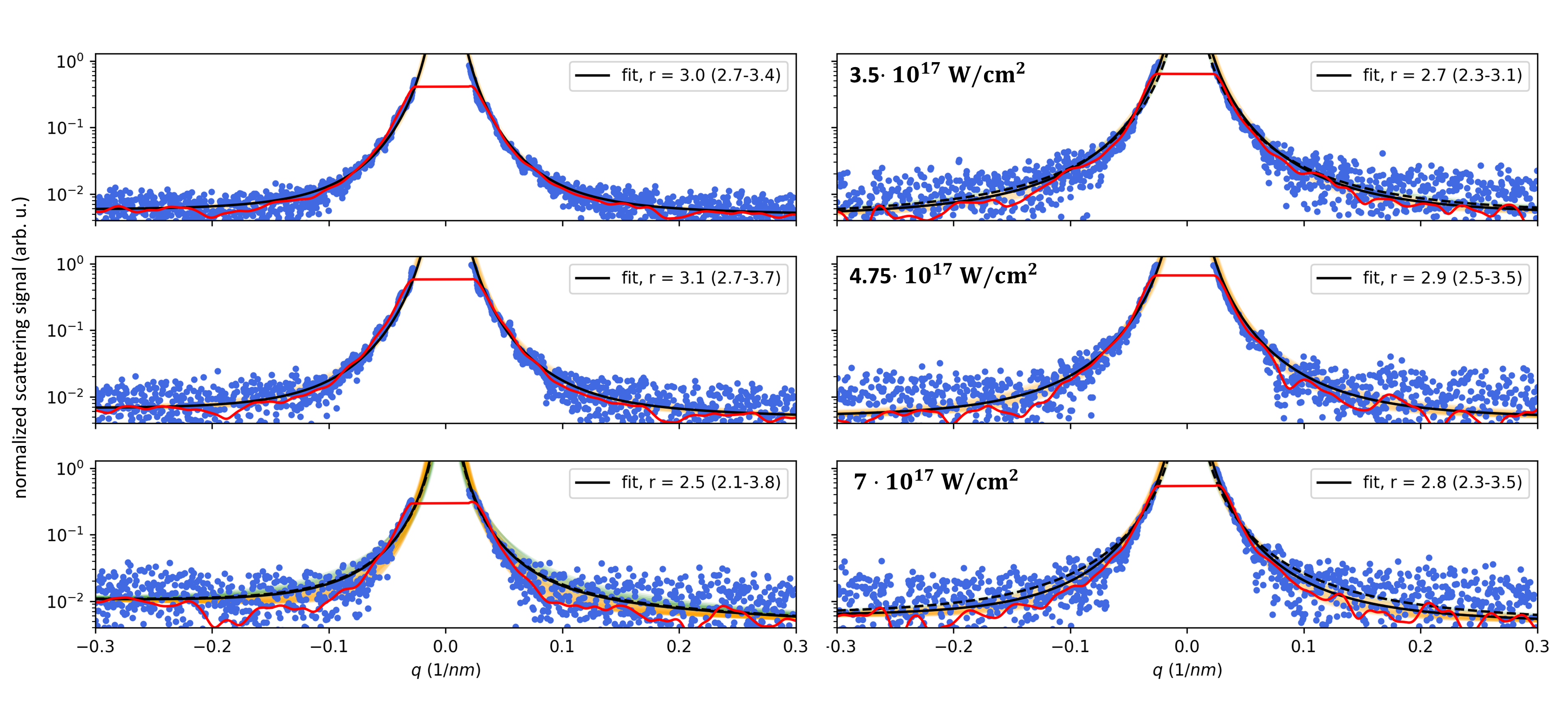}
\caption{Same as \ref{fig:delay_scan_profiles} but for the streak profiles of the scattering patterns shown in Fig.~\ref{fig:I_scan}.}
  \label{fig:I_scan_profiles}
\end{figure*}
We start the analysis of the streaks by comparing their lengths, i.e the intensity fall-off with scattering angle (i.e. $I(q)$), cf. Figs.~\ref{fig:delay_scan_profiles} and~\ref{fig:I_scan_profiles}. 
In none of the shots a Debye-Waller like exponential roll-off at large $q$-values could be observed. 
%As we are limited to a q-range less than $q=0.1/\mathrm{nm}$ due to the low signal strength (i.e. a $\sigma$ of more than $50\unit{nm}$, see below) we conclude that the scattering surfaces are always sharper than the detectable minimum value within the dynamic range. 
Since the q-range with signal above the background is limited in most of the shots to $q < q_{max}\cong0.1/\unit{nm}$, this means that the surface roughness would be less than
\begin{equation}
    \sigma_{max} \approx \frac{1}{q_{max}}\approx 10\unit{nm}. 
\end{equation}
In reality this value is even much larger since photon number Poisson statistics, background uncertainty, and the uncertainty of $r$ and its fit correlation with $\sigma$ add considerable fit uncertainty (the latter alone is approx. a factor of 2). 
This means that in our case due to the low dynamic range limiting us to very small q-value, the Debye-Waller factor cannot be discriminated from unity in all shots. 
Setting it to $1$ in Eqn.~\eqref{eqn:Debye}, the scattering intensity can thus simply be written as
\begin{equation}
    I(q) = \frac{a}{q^r} + b
    \label{eqn:intensity}
\end{equation}
where $a$ is the proportionality factor, $b$ comprises the radiation and detector background signal and $r$ should take on values between $2$ and $3$. 
%The exponent $r$ is characteristic for the shape of the target. 
%For example, for the XFEL projection through a step-like planar plane, $r=2$; for the projection through a circular wire vertically through the z-axis, $r=3$. 
We fitted all the streaks with with~Eqn.~\eqref{eqn:intensity} and find a value close to $r=3$ for all cases, see %, as confirmed within the errorbars by the profiles shown in 
Fig.~\ref{fig:delay_scan_profiles}. % except the filamented cases where the selection of a single streak is difficult and the fit therefore questionable. 
This means that in all cases the shape of the scattering front remains that of a cylinder and is not significantly compressed by the laser towards flatter planar geometry. 
Having found no evidence for expansion nor compression, we conclude that the visual impression of shorter or longer streaks is consistent with being only due to different signal levels instead of a change of $r$ or $\sigma$. \\

\begin{figure*}
\centering
  \includegraphics[width=\linewidth]{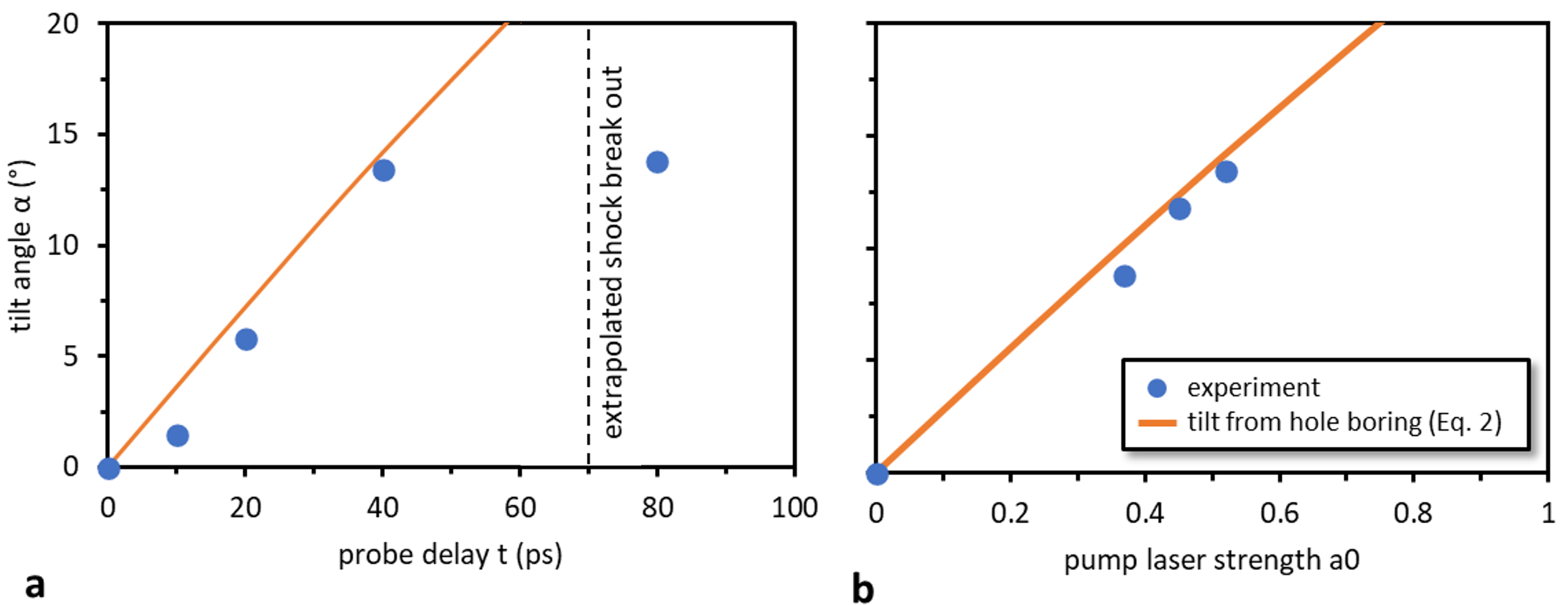}
\caption{Tilt angle of the streak as a function of probe delay at full pump laser intensity (a) and as a function of pump laser intensity at $40\unit{ps}$ probe delay (b) (black points), compared to a surface bending forward proportionally to the hole boring velocity computed with the local laser intensity following its transverse Gaussian shape (orange lines). }
  \label{fig:angles}
\end{figure*}

Secondly, we analyze the tilt angle of the streak in the main-shot relative to the pre-shot. 
Fig.~\ref{fig:angles} shows the tilt angles as a function of delay for full laser intensity (a), and as function of laser intensity at 40 ps delay (b). 
As described above, there are several mechanisms that in principle can cause a time-varying signal orientation. 
Some we can exclude based on our measurements by realizing that our XFEL spatial overlap was vertically shifted approx. $20\mum$ above the laser interaction point (i.e. $>2$ FWHM of the XFEL pulse spot size). 
This is why we only see one streak - the other streak in the opposite direction is simply not in the XFEL field of view. 
However, this allows us to infer the direction of the tilt of the scattering surface. 
In the present case, we conclude that the tilt must be in laser forward direction. 
This means that those processes that cause a tilted density contour tilt towards the laser can be excluded, i.e. front surface plasma expansion and rear surface rarefaction wave. 

We are left with the possibilities of rear surface tilt of the target due to plasma expansion into vacuum, or a tilt of the front surface by hole boring or a compression (shock) front traveling forward, cf. Fig.~\ref{fig:setup:angles_explained} (orange lines). 
To connect the tilt angle of the streak with the plasma electron density contour more quantitative, we adopt a very simple model. 
We assume that the scattering surface is moving forward with a velocity that depends on the local laser intensity. 
In particular, the surface longitudinal position as a function of the transverse offset shall be Gaussian with the same FWHM as the laser amplitude focal spot size $w_{y}$. 
Then it can easily be geometrically derived that the target position $A$ along the target normal is connected to the angle $\alpha$ of the normal at the point of inflection (i.e. the tilt angle) via
\begin{equation}
    \alpha = \sqrt{e}\frac{w_{y}}{2\sqrt{\mathrm{ln}2}A}
    \label{eqn:model}
\end{equation}
From the delay scan of $\alpha$ in Fig.~\ref{fig:angles}a we can then directly obtain the peak forward velocity of $v_f=A/\Delta t\cong 0.1\pm \unit{\mum/ps}$ until a probe delay of $\Delta t \gtrsim 40\unit{ps}$. 
Obviously at $80\unit{ps}$ delay the measured tilt does not increase anymore. 
However, extrapolating the forward velocity, we would expect the scattering front to exit the rear of the wire at $\approx 70\unit{ps}$ if launched from the front at $t=0$ (However, one needs to be careful with the interpretation, since we present only single shot results). \\ 
From the intensity scan in panel (b) we infer that the forward velocity is proportional to the laser strength parameter $a_0$, i.e. the square root of the laser intensity, $v_f = c_f a_0$, with a proportionality constant of $c_f\cong 0.19\pm \unit{\mum/ps}$. \\
The forward velocity extracted from both the delay scan and the intensity scan is in remarkable agreement with the hole boring velocity from Eqn.\eqref{eqn:v_HB}, $v_{HB}\cong 0.2 \frac{\mum}{\mathrm{ps}}\cdot a_0$, 
%\begin{equation}
%    v_{HB} = a_0 \sqrt{\frac{1}{2}\frac{n_c}{n_e}\frac{m_e}{m_p}}\cong 0.2 \frac{\mum}{\mathrm{ps}}\cdot a_0,
%    \label{eqn:v_HB}
%\end{equation}
and for the full intensity case agrees with the compression front velocity in the aforementioned simulation.

Finally, we point out that the tilted streak transitions into a set of multiple streaks at full intensity after $20\unit{ps}$. 
This is indicative for a modulation of the scattering surface. 
This filamentation starts around $20\unit{ps}$, is most visible at  $40\unit{ps}$ and is not visible at $80\unit{ps}$ after the laser irradiation as strong anymore, though there is still a faint second streak visible. 
Potentially the instability is still present, but its spatial frequency decreased so that the field of view of the XFEL comprises less filaments. 
The growth of the instability apparently continues, as the angular spread between the streaks at larger delay values is larger than at earlier times, i.e. the modulation depth seems to be increasing with time. 
In  most shots we additionally see remnants of the streak in horizontal direction, perpendicular to the undriven wire surface. 
We believe that this is primarily due to scattering of the outer wings of the XFEL spot at the rear surface or vertical distant, unperturbed regions.

\section*{Conclusions}
In the experiment presented above we have employed SAXS to measure the target solid density response upon HI laser irradiation. 
We interpret our measurements with scattering at a compression front inside the target after it detached from the laser radiation pressure accelerated front surface. 
This follows from the following arguments: 
First, the simulation in \cite{Gaus2021} that predict a compression shock front persisting over 10s of ps. 
Secondly, the front and rear surface are expected to quickly smooth out due to plasma expansion, so that the streak cannot originate from the ablated wire surfaces. 
The tilt angle orientation of the measured SAXS streaks that indicates a forward deformation of the scattering surface, i.e. we can exclude the rear surface rarefaction wave. 
Rather, it is in agreement with a laser compression front following the lateral laser intensity shape. 
Moreover, the change of the tilt angle as a function of probe delay and pump intensity is in agreement with the hole boring velocity scaling. 
Surprisingly, apparently the scattering surface remains quite sharp over many picoseconds, since the observed streak lengths did not change. 

At full laser intensity we observe the streak to filament at a probe time $20\unit{ps}$ after HI irradiation. 
The number of visible streaks then decreases and the distance between streaks seems to increase between $20\unit{ps}$ to $80\unit{ps}$. 
This indicates that if the signal was due to a sinusoidal-like modulation of the scattering surface, its period and amplitude would grow with time. 
With the limited data available from this small study, we cannot specify which instability type causes the splitting of the streak. 
However, apparently similar modulations on similar hydrodynamic time scales have been observed before by means of optical probes and hence at different spatial resolution and lower plasma density\cite{martin}. 

In this paper we reported on the plasma reaction after \emph{near-relativistic} pump laser irradiation of a thin wire. 
This study is limited to only a few shots, which means that we cannot give an estimate of the reproducibility.  
Additionally, the fits of the streak to obtain $r$ are connected with large uncertainties and the fit of the surface expansion was not possible at all since the exponential roll-off due to the Debye-Waller factor was not visible (likely hidden in the background). \\
While absorption and phase contrast x-ray imaging~\cite{Schropp2015} could also have been employed to visualize the dynamics of the bulk density on those scales, the spatial resolution of these methods is still limited to a few $100\,\rm nm$ and is thus larger than the relevant scales during or shortly after the laser pump irradiation itself -- which is our future aim in order to eventually probe directly the early dynamics. 
For \emph{ultra-relativistic} intensities, as they are relevant for many important applications such as ion acceleration, isochoric heating, or high harmonic generation, much faster and higher resolving probing is currently underdevelopment. 
The next generation of high intensity XFEL experiments are currently starting at European XFEL, using shorter probe pulses, larger detector distance and dynamic range, and background radiation suppression with chromatic mirrors. 
%Additionally, in order to increase the minimum tilt angle and to increase the streak signal intensity, the scattering streak should be as narrow as possible. 
%This means the wire shall be transversely as flat as possible without much surface roughness, as was demonstrated by our group using wires etched from silicon wavers\cite{Kluge2017}. 
Then, the onset and early dynamics of the filamentation and plasma expansion could be studied and directly compared to existing models and particle-in-cell simulations. \\
%Outlook: 2019 MIST scattering image after 1ps: showing already splitting.

\section*{Data availability}
The data that support the findings of this study will be openly available once the peer-reviewed article is published. 

\begin{acknowledgments}
This work was partially supported by DOE Office of Science, Fusion Energy Science under FWP 100182. Use of the Linac Coherent Light Source (LCLS), SLAC National Accelerator Laboratory, is supported by the U.S. Department of Energy, Office of Science, Office of Basic Energy Sciences under Contract No. DE-AC02-76SF00515. The experiments were performed at the Matter at Extreme Conditions (MEC) instrument of LCLS, supported by the DOE Office of Science, Fusion Energy Science under contract No. SF00515. 
Christian Gutt acknoledges funding by DFG (GU 535/6-1). 
This work has also been supported by HIBEF (www.hibef.eu) and partially by H2020 Laserlab Europe V (PRISES) contract no. 871124, and by the German Federal Ministry of Education and Research (BMBF) under contract number 03Z1O511. 
\end{acknowledgments}

\bibliography{Mendeley}

\end{document}